# SERVICE ORIENTED ARCHITECTURE A REVOLUTION FOR COMPREHENSIVE WEB BASED PROJECT MANAGEMENT SOFTWARE

A.R. Khan[1], Rquaiya Khan[2], Trimbak R Sontakke[3], Shraddha R Khonde[4], Revati Wahul[5], Mahtab alam[6]

[1,4,5]M.E.S College of Engineering, Pune -411001
[1]ahmed.khan@mescoepune.org
[4]shraddha.khonde@mescoepune.org
[5]rmwahul@mescoepune.org

[2]Uttrakhand University, Uk.
ruqaiyamaqsood@gmail.com

[3]Siddhant College of Engineering.
trsontakke@gmail.com

[3]Poona College, Pune.
mahtabalam@hotmail.com

## ABSTRACT

*Service Oriented Architecture A Revolution for Project Management Software* has changed the way projects today are moving on the fly with the help of web services booming the industry. Service oriented architecture improves performance and the communication between the distributed and remote teams. Web Services to Provide Project Management software the visibility and control of the application development lifecycle-giving a better control over the entire development process, from the management stage through development. The goal of Service Oriented Architecture for Project Management Software is to produce a product that is delivered on time, within the allocated budget, and with the capabilities expected by the customer. Web Services in Project management Project management software is basically a properly managed project and has a clear, communicated, and managed set of goals and objectives, whose progress is quantifiable and controlled. Resources are used effectively and efficiently to produce the desired product. With the help of service oriented architecture we can move into the future without abandoning the past. A project usually has a communicated set of processes that cover the daily activities of the project, forming the project framework. As a result, every team member understands their roles, responsibilities and how they fit into the big picture thus promoting the efficient use of resources.

# 1. INTRODUCTION

No other industry has such a degree of failure in its projects as the IT industry. No wonder, really: Designing databases is more of an art than a craft, and the same goes for developing software. Our industry is very much based on individual skills, and as such, results are often unpredictable. That's why it is said that 50% of all IT projects are failures. So, software project management is crucial in order to manage, control and disclose early signs of project failure, so they can be corrected, and the project brought back on the right course again. This research work aims at implementing web services for project management software. As the number of projects are increasing day in and out project management using traditional method is becoming complicated and the resources that are required for managing the projects are shrinking so we need a project management system that is scalable and can manage multiple projects simultaneously and with good throughput. CPMS is cloud based project management system. It consists of 3 roles customer, developer and project manager. Here one person can have many roles and each role have different access rights. Customer can send his requirements to project manager. Project manager create project and add developers to it. Project manager do scheduling of project and assign tasks to developers. Developers build different modules based on customer requirements and scheduling done by project manager. After completion of project manager deploy it at customer side and take its feedback. In this process developers can follow and unfollow projects. Other modules in CPMS are schedulers, add project module, invite people module, risk management template, discussion platform, COCOMO calculator. There are two schedulers in CPMS personal scheduler and common scheduler. Using personal scheduler, developer can check his tasks project wise and using common scheduler he can see other's tasks. Only project manager can add and assign tasks.Project management is the science (and art) of organizing the components of a project, whether the project is development of a new product, the launch of a new service. A project is not something that's part of normal business operations. It's typically created once, it's temporary, and it's specific. As one expert notes, "It has a beginning and an end." A project consumes resources (whether people, cash, materials, or time), and it has funding limits.Project management has been practiced for thousands of years dating back to the Egyptian epoch, but it was in the mid-1950 that organizations commenced applying formal project management tools and techniques to complex projects. During the 1960s and 1970s, PERT and CPM increased their popularity within the private and public sectors. The use of project management techniques in the 1980s was facilitated with the advent of the personal computer and associated low cost project management software. Hence, during this period, the manufacturing and software development sectors commenced to adopt and implement sophisticated project management practices as well. By the 1990s, project management theories, tools and techniques were widely received by different industries and organisations.

"The importance of Project Management" is an important topic because all organisations, whether small or large, at one time or other, are involved in implementing new undertakings. These undertakings may be diverse, such as, the development of a new product or service; the establishment of a new production line in a manufacturing enterprise; a public relations promotion campaign; or a major building programmed. Whilst the 1980's were about quality and the 1990's were all about globalisation, the 2000's are about velocity. That is, to keep ahead of their competitors, organizations are continually faced with the development of complex products, services and processes with very short time-to-market windows combined with the need for cross-functional expertise. In this scenario, project management becomes a very important and powerful tool in the hands of organizations that understand its use and have the competencies to apply it.

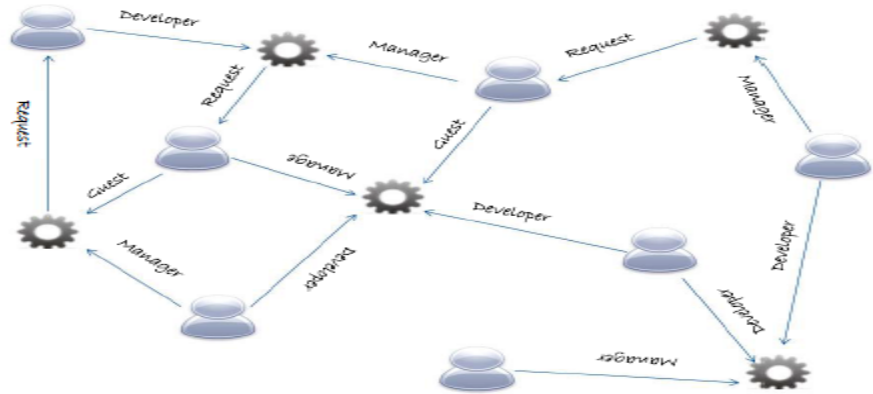

Fig 1: Login with multiple roles.

## 2. SYSTEM ARCHITECTURE

Basically our system is divided in to two parts
    1]. Web Role

    2]. Worker Role

Explanation of these roles is as follow.
    **1]. Web Role:**

    Web role basically used for creating user interface which can be view to user. We are using MVC 3 architecture in web role.
    So Web role is divided into three parts
        a. **View:**
        This part creates user interface for the client. User's request can be capture from this view. After getting user request it invokes appropriate controller function which gives proper response to request.

        b. **Controller:**
        All requests for new view is first goes to controller which ultimately transfer control to respected view. All business logic is implemented in controller. In our system all cloud service for respected request is called from controller.

        c. **Model:**
        Model basically contains all class, which are used to transfer data from controller to view.

    **2]. Worker Role**

Worker role basically perform all business logic so that it can be used by many web role. It access database which are required in our project. Web role knows to worker role by using its endpoint only by which we can access worker role.

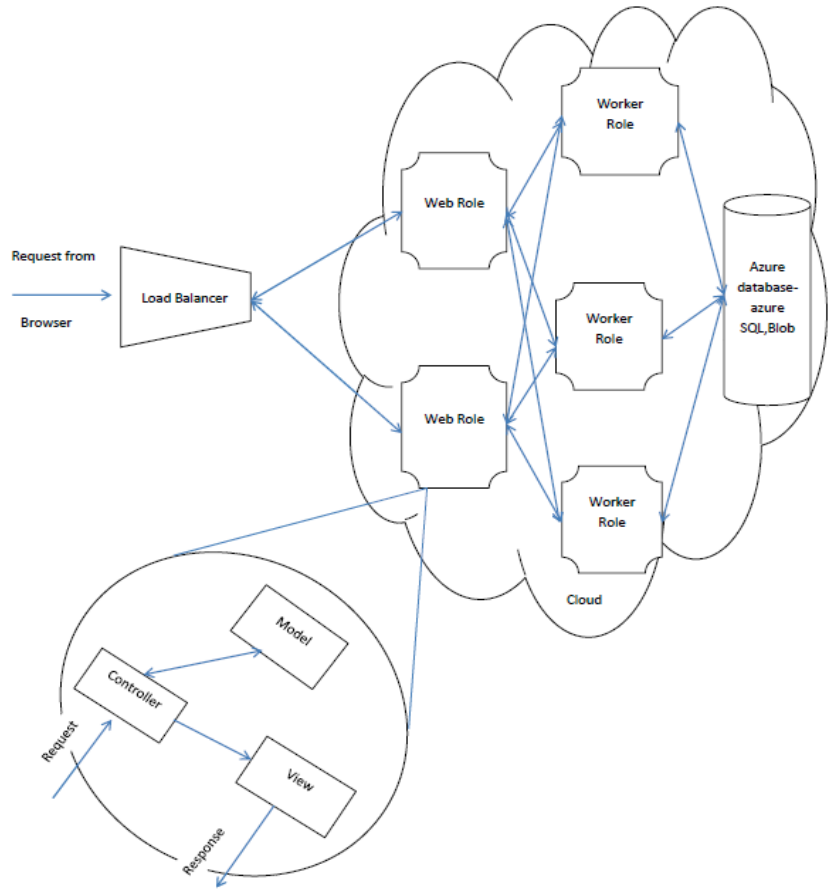

Figure 2. Architecture of the System

## 3. WORKING EXPLANATION

CPMS will be deployed on Microsoft azure which provide azure SQL server which is quite similar to normal SQL server which we use on local machine. SQL Azure is a highly available and scalable cloud database service built on SQL Server technologies. With SQL Azure, developers don't have to install setup or manage any database. High availability and fault tolerance is built-in and no physical administration is required .We normalize our database table such that we try to decrease query execution time as well as reduce data redundancy considerably. We are also using blob storage which is specially design to store files and image to store files which will be uploaded by userProject Management software can be deployed on to the IIS server and different persons sitting on there matching can access this application as PMS is a web based software it contains different sections that can be used for managing and mentioning the software project. Basically this software contains the new registration screen that can be used to register for using the functionalities and features of this system after the user or the client is registered he can log into the system but can not access the different functionalities of the software unless the administrator provides him the rights to access the different functionalities of the software system. Once the client has registered successfully a message will be send to the administrator who will provide rights to the client to access the different parts of the project the administrator will give rights to the client to access the screens that the client is supposed to see and access. The employee of the company can also register with PMS while employee registration the employee has to specify his designation. The PMS allows you to manage the entire software

## 4. DESIGN OF THE SOFTWARE

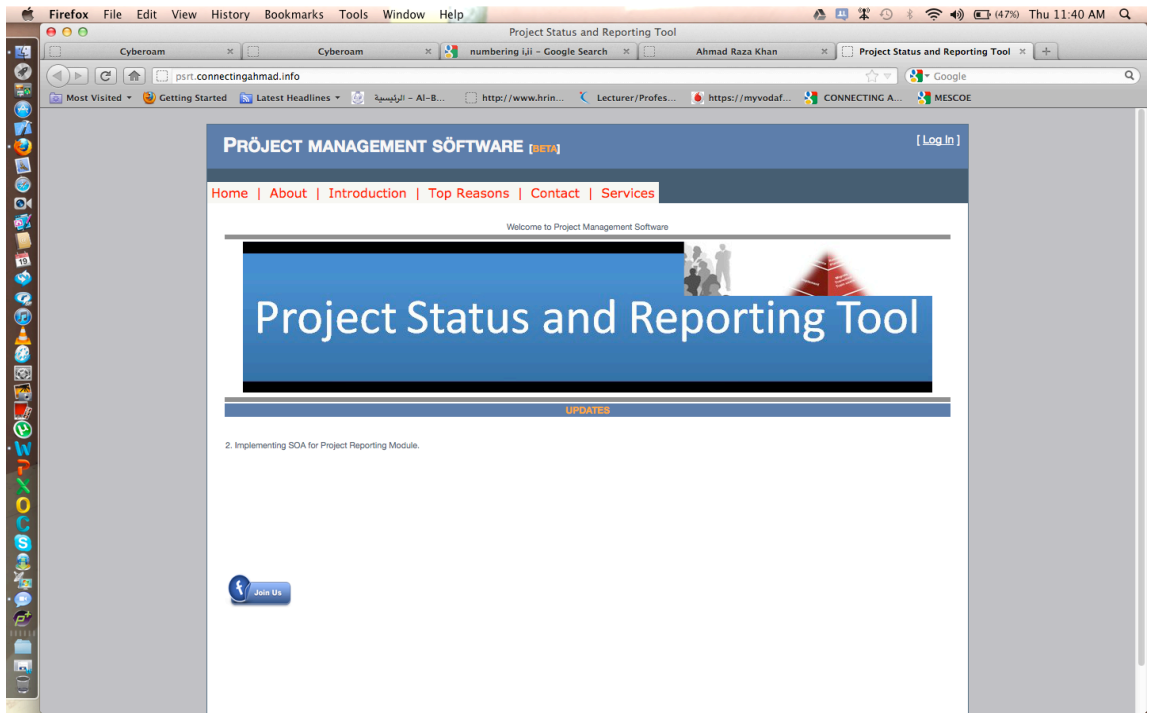

Fig 3. Login Screen of Project Management

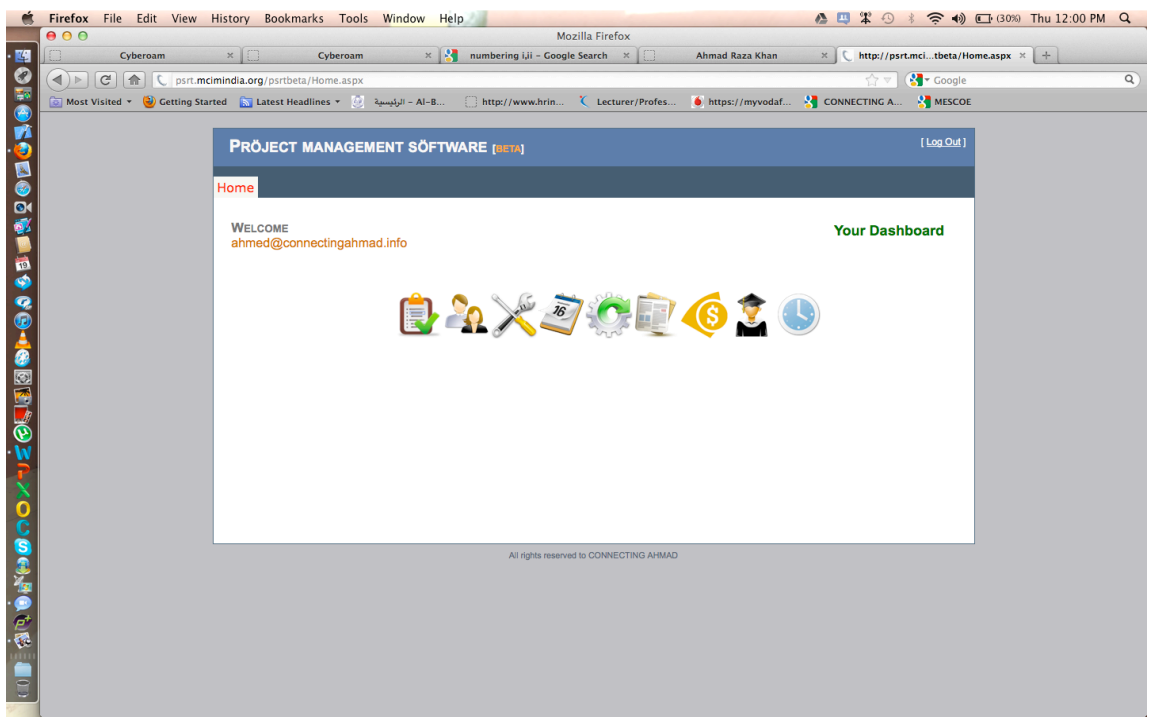

Fig 4. User interface for dashboard.

## 5. CONCLUSION

Basic idea of our application is avoid the software project failure due to lack of user input, incomplete requirements & specifications , changing requirements & specifications , unclear objectives of project which usually occurs because of lack of communication between client and remotely situated software development teams. Our application provides an integrated environment from managing change and configuration requirements and a system that is designed to give you visibility into the development process-one that will help you maintain control of the application lifecycle .Our application will provide platform for interaction between client and remotely situated teams as well as all stake holder from project can monitor project development procedure. Cloud based PMS will definitely increase productivity and success rate of software development. And important feature is our application will be SAAS. So user can use these services according to their requirement.

## ACKNOWLEDGEMENTS

I would like to thank "Ruqaiya Ahmad Raza Khan" for constantly helping me and guiding me for the preparation of this paper and my research work. I would also like to thank "Prof. Dr. Trimbak R Sontakke" for giving me this opportunity to present this paper at international level and providing me with the necessary resources as and when required.

**Author**

Prof. Mr. Ahmad Raza Khan,

Assistant Professor,

M.E.S College of Engineering

Pune- 411001.

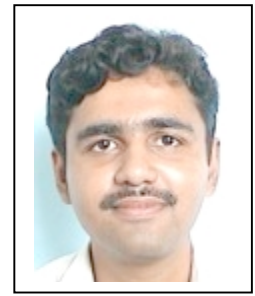

**Co-Authors**

Mrs. Ruqaiya Ahmad Raza Khan,

Student,

Uttrakhan University

Nanital.

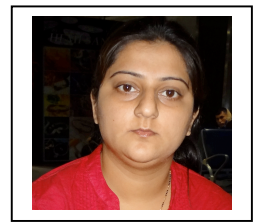

**Co-Authors**

Porf. Dr. Trimbak R Sontakke,

Principal,

Siddhant College of Engineering,

Sudumbare Pune.

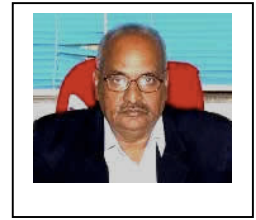

**Co-Authors**

Porf. Mrs.Shraddha R Khonde,

Assistant Professor,

M.E.S College of Engineering,

Pune- 411001.

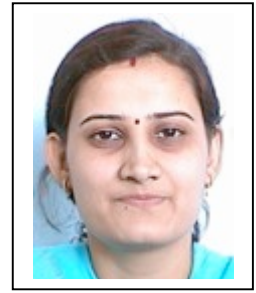

**Co-Authors**

Porf. Ms.Revati Wahul,

Assistant Professor,

M.E.S College of Engineering,

Pune- 411001.

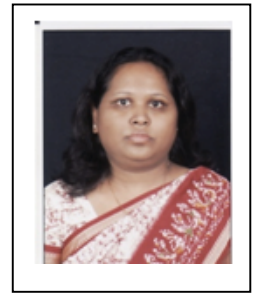

**Co-Authors**

Porf. Mahtab Alam,

H.O.D,

Poona College,

Pune- 411001.

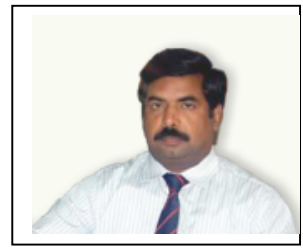